\documentclass[aps,showpacs,graphicx]{revtex4} 
\usepackage{amssymb} 
\usepackage[dvips]{graphicx} 
\usepackage[english]{babel} 
\usepackage{indentfirst} 
\usepackage{amsxtra} 
\usepackage{amsmath} 
\usepackage{supertabular} 
\usepackage{multirow} 
\usepackage [mathcal]{eucal}
\usepackage {float}  
\usepackage{lscape} 
\usepackage{rotating}
\newcommand{\beq}{\begin{equation}} 
\newcommand{\eeq}{\end{equation}} 
\newcommand{\beqn}{\begin{eqnarray}} 
\newcommand{\eeqn}{\end{eqnarray}} 
\newcommand{\br}{{\bf r}}

\newcommand{\vamf}{{ (v_\alpha^{\rm MF})^2}} 
 
\newcommand{\vabcs}{{ (v_\alpha^{\rm BCS})^2}} 
\newcommand{\vahfb}{{ (v_\alpha^{\rm HFB})^2}} 
\newcommand{\rbox}{{R_{\rm box}}} 
\newcommand{\emax}{{ \epsilon_{\rm max}}} 
\usepackage[usenames]{color} 
\usepackage{ulem}

\begin{document} 
 
\noindent 
\title{Neutron gas and pairing} 
 
\author{ M. Anguiano$^{\,1}$, A. M. Lallena$^{\,1}$,
R. Bernard$^{\,2,3}$,  G. Co'$^{\,4,5}$}
\affiliation{
$^1$ Departamento de F\'\i sica At\'omica, Molecular y 
  Nuclear, Universidad de Granada, E-18071 Granada, Spain \\
$^2$ \'Ecole normale sup\'erieure Paris-Saclay, 61 Avenue du Pr\'esident Wilson, F-94230 Cachan, France\\
$^3$ CEA, DAM Ile-de-France, F-91297 Arpajon, France\\
$^4$ Dipartimento di Matematica e Fisica ``E. De Giorgi'', 
  Universit\`a del Salento, I-73100 Lecce, Italy \\ 
$^5$ INFN Sezione di Lecce, Via Arnesano, I-73100 Lecce, Italy \\ 
}  

\date{\today}

\bigskip 
 
\begin{abstract} 
We study the emergence of neutron gas effects in
the description of nuclei with large neutron excess
within the Bardeen-Cooper-Schrieffer approach. 
We consider Ni and Sn isotopes where, 
in the literature, these effects have been found.
We investigate the role of the single particle states 
with positive energy generating the neutron gas,
and we find that the contribution of these states is numerically 
irrelevant for the various observables that we evaluate. 
\end{abstract} 

\maketitle 

\section{Introduction} 
\label{sec:intro} 

The study of exotic nuclei close to the neutron-drip line has generated 
great interest in these last 
years. For these systems, uncommon nuclear properties have been predicted
and some of them have been experimentally identified \cite{gad08,sor08,wie13,ste13}. 

The theoretical description of these nuclei with a large neutron excess requires an accurate 
treatment of pairing correlations. Among the various approaches built to handle these correlations,
one of the most used is that based on the  Bardeen-Cooper-Schrieffer (BCS) theory  
\cite{boh58,row70,rin80,suh07}. 
In this approach, the nucleon-nucleon pairing interaction 
modifies the occupation probabilities of a set of single particle (s.p.) states 
generated within a mean-field (MF) model.

Despite the fact that the MF+BCS approach is widely utilized
to study nuclei throughout the whole nuclear chart, 
some questions have been raised on its reliability 
in the description of neutron rich nuclei. These doubts, in the literature,
have been summarized under the name of {\sl neutron, or particle, gas problem} 
\cite{dob84,dob96,dob96b,dob99}.

To clarify the origin and the details of this problem, let us 
consider the one-body density matrix (OBDM) 
defined as \cite{ari07} 
\beq
\rho(\br,\br') \,=\, \frac {A} {\langle \Psi | \Psi \rangle}
\int {\rm d}^3 r_2 \, \cdots {\rm d}^3 r_A \,
\Psi^*(\br, \br_2, \cdots ,\br_A) \, \Psi(\br', \br_2, \cdots ,\br_A) 
\, ,
\label{eq:denmat}
\eeq
where we have indicated with $\Psi$ the eigenstates of the hamiltonian 
describing a finite system of $A$ interacting nucleons. 
In the ground state of a bound nucleus, 
the nucleons are localized in space, and, therefore, the OBDM 
verifies the conditions
\beq
\lim_{\br \rightarrow \infty} \rho(\br,\br') \,=\, 
\lim_{\br' \rightarrow \infty} \rho(\br,\br') \,= \, 0 
\label{eq:limit}
\, .
\eeq

In the MF approach the many-body hamiltonian is 
simplified: it can be written 
as a sum of non-interacting s.p.
hamiltonians and the many-body wave functions 
can be expressed as Slater determinants 
of eigenstates $\phi_\alpha(\br)$ 
of these s.p. hamiltonians. In this case, the 
OBDM assumes the form:
\beq
\rho^{\rm MF}(\br,\br') \, = \, \sum_\alpha \, \vamf \,
\phi^*_\alpha(\br)  \, \phi_\alpha(\br') \, \Theta(\epsilon_{\rm F} -\epsilon_\alpha)
\, ,
\label{eq:obdmhf}
\eeq 
where we have indicated with
$\epsilon_{\rm F}$ the Fermi energy, 
with $\vamf$ and $\epsilon_\alpha$, respectively,
the occupation probability and the energy of the s.p. state  
$\phi_\alpha(\br)$, and with $\Theta(\alpha)$ the step function. 
In a MF model all the states up to the Fermi surface are fully
occupied, $\vamf=1$ if $\epsilon_\alpha \leq \epsilon_{\rm F}$, and those above it are completely empty,
$\vamf=0$ if $\epsilon_\alpha > \epsilon_{\rm F}$.

If the set of s.p. states is obtained in Hartree-Fock (HF) calculations, a variational 
procedure that consists in finding the energy minimum by modifying the s.p. wave functions is performed. 
The values of the occupation probabilities remain unaltered. 
Since the $\phi_\alpha$ states below the Fermi surface, {\it i.e.} those 
involved in the sum in Eq. (\ref{eq:obdmhf}), are bound, they 
are localized in space and, consequently, the boundary conditions 
of Eq. (\ref{eq:limit}) are still satisfied.

In the BCS approach, the pairing acts on the MF picture by changing
the values of the occupation probabilities, while the s.p. wave functions
are not modified. 
The OBDM is given by the expression
\beq
\rho^{\rm BCS}(\br,\br')\, = \, \sum_\alpha \, \vabcs \,
\phi^*_\alpha(\br) \,  \phi_\alpha(\br')
\, .
\label{eq:obdmbcs}
\eeq 
It is worth noting that, at variance with the MF
expression (\ref{eq:obdmhf}), in Eq. (\ref{eq:obdmbcs}) the sum
does not have an upper limit. 
This means that, in principle, all the s.p. wave functions contribute 
to $\rho^{\rm BCS}$, even those in the continuum which have an 
oscillating behavior at the boundaries. Under these circumstances, 
compliance with the conditions (\ref{eq:limit}) 
cannot be guaranteed for $\rho^{\rm BCS}$ and this is the source of the neutron gas problem.

This problem is absent in Hartree-Fock-Bogoliubov (HFB) theory, the other 
widely used approach which treats pairing correlations \cite{rin80}. 
In this case, the variational principle is applied by changing
both s.p. wave functions and occupation probabilities at once. 
The OBDM obtained in the HFB theory can be written as 
\beq
\rho^{\rm HFB}(\br,\br')\, = \, \sum_\alpha \, \vahfb \,
\zeta^*_\alpha(\br) \,  \zeta_\alpha(\br')
\, ,
\label{eq:obdmhfb}
\eeq 
formally equivalent to the BCS density given in Eq. (\ref{eq:obdmbcs}).
However, in this case, the wave functions $\zeta_\alpha(\br)$ belong to the 
so-called ``canonical basis'' which, by construction, are limited in 
space. For this reason  in HFB calculations 
the boundary properties of Eq. (\ref{eq:limit}) are always fulfilled.

Summarizing, we can say that 
the BCS theory considers the presence of free nucleons even in the
nuclear ground state. In Refs. \cite{dob84,dob96,dob96b,dob99} it
has been pointed out that this feature, in neutron rich nuclei,
affects the values of the neutron distribution radii,
which appear to be too large with respect to those found in HFB calculations, 
and also the asymptotic behavior of the OBDMs showing long distance tails
abnormally high. These facts have been considered as evidences of the neutron gas problem.

Even though these considerations would induce us to abandon 
the MF+BCS approach in favour of the HFB one, there are some
nice features of the former approach that make it still widely
utilized. It is easier to solve the BCS equations than those 
of the HFB theory, the physical interpretation of the results is more 
direct and the many-body wave functions are more handily usable 
to calculate nuclear excited states as, for example, within the 
Quasi-Particle Random Phase Approximation (QRPA) theory. 

Furthermore, in our experience, we used the MF+BCS approach 
to study various nuclei in different regions of the nuclear chart
\cite{ang14,ang15,ang16a,don17} and we never encountered 
problems related to the presence of neutron gas. 
This result was quite astonishing, and induced us to study the 
quantitative relevance of this problem in order to test the 
reliability of the MF+BCS  approach and define the limits of validity 
in its application. 

We have carried out our investigation by comparing results obtained 
with both MF+BCS and HFB approaches by using various types
of nucleon-nucleon and pairing interactions. We present here
the results that we have obtained for those isotopes considered in Refs. 
\cite{dob84,dob96,dob96b,dob99} where the neutron gas problem
was identified.

Since the quantitative relevance of the neutron gas effects is the
subject of our investigation, we discuss with some detail
in Sec. \ref{sec:theory} those aspects of our calculations related
to their numerical stability. In Sec. \ref{sec:results},
we compare our results to those found in
the literature \cite{dob84,dob96,dob96b,dob99}.
First we consider neutron radii and density distributions in Ni and
Sn isotopes. To verify the quantitative effects of the neutron gas
in other observables, we present the 
proton elastic scattering cross sections calculated for the $^{150}$Sn nucleus.
We summarize our results in Sec. \ref{sec:conclusions} and we conclude
that, from the quantitative point of view, the neutron gas problem is 
irrelevant and MF+BCS calculations are reliable in all the regions of the
nuclear chart table.

\section{Details of the calculations}
\label{sec:theory}

The aim of our study is the evaluation of the quantitative
relevance of the effects induced by the neutron gas in MF+BCS
calculations. For this reason, the numerical reliability of the 
computational methodologies adopted to solve the various 
equations defined by the model is an essential ingredient of
our investigation. Therefore, in this section, 
we present in detail all the key ingredients of our calculations 
and we discuss the implications of the various inputs either
of physics or numerical type.

In the present work, we have considered only spherical nuclei. For each of them a MF
s.p. basis has been generated by solving HF equations. 
By exploiting the spherical symmetry of the system, 
the HF equations have been
expressed as integro-differential non-linear equations 
depending only on a single variable, the distance from the center of coordinates, $r$. 
We have defined a maximum value, $\rbox$, of this distance, and in this position
we have imposed infinite well boundary conditions. The value of $\rbox$ is 
one of the numerical inputs of our approach and depends on the size of 
the nucleus investigated. 

The HF equations are solved iteratively, by considering only the s.p. wave functions 
with s.p. energies $\epsilon_\alpha \leq \epsilon_{\rm F}$. The numerical solution of these equations 
is obtained by using the plane wave expansion technique developed 
in Refs. \cite{gua82a,gua82b}. 
The number of plane waves considered in the expansion 
is equal to the number of mesh points employed to describe the $r$ space up to $\rbox$.
This is the second, numerical, input of our approach and we have found that a mesh size of
the order of $0.1\,$fm is enough to ensure the numerical stability of the HF results
for the nuclei studied. This technique allows a treatment of the non-local
Fock-Dirac term of the HF equations without any approximation. 

In our procedure, when the minimum of the binding energy 
is found, and for this purpose we have chosen
a convergence value of $1\,$eV, the Hartree and the Fock-Dirac
potential terms \cite{rin80}, built with the s.p. wave functions 
with $\epsilon_\alpha \leq \epsilon_{\rm F}$,
are used to calculate also the s.p. wave functions 
and energies of the states above the Fermi surface. 
Because of the infinite well boundary conditions, 
the s.p. spectra are discrete, even for energies larger than zero. 
While the bound s.p. states are extremely stable against 
the choice of the $\rbox$ value
(we observe a change in the value
of the s.p. energy of about one part on a million for a 20\% change of $\rbox$), 
the s.p. states with positive energies are very sensitive to it. 
It is worth pointing out that this affects all the theories making use
of the full set of HF, or more in general MF, s.p. states
because all of them have to deal with the ``discretized continuum'' 
part of the s.p. spectrum. 

In our approach, the BCS equations are solved by means of
an iterative procedure and by using standard numerical methods \cite{suh07}.
The configuration space we have considered is composed by the set of discrete s.p. wave functions,
with both negative and positive energies, generated by the previously described HF calculations. 
The value of the box size required for the integrations in $r$-space 
is the same as that employed in the HF calculations. The only new input 
is the maximum value of the energy of the s.p.
states considered, $\emax$, 
which determines the size of the configuration space.

In the present study, the effective nucleon-nucleon interaction that
we have considered is a finite-range force of Gogny type, specifically
in its D1S parameterization. This is the most traditional,
and widely used, Gogny force, and its parameters have been chosen
to fit experimental values of binding energies and charge root mean
square (rms) radii belonging to a large body of nuclei in all  the 
regions of the nuclear chart \cite{ber91}. We have consistently used
this interaction in both steps of our calculations, HF and BCS.
As is it clearly discussed in Ref. \cite{dec80}, the finite-range feature of 
this  interaction induces a natural cut in the coupling to high energy s.p. 
states due to pairing. This avoids to insert in the theory additional physics 
inputs, such as a specific pairing interaction related to the size of the
s.p. configuration space \cite{dob84,dob96,dob96b}. 

\begin{table}[htb] 
\begin{center} 
\begin{tabular}{ccccccc}
\hline\hline
&~~~&  &~~~&\multicolumn{3}{c}{$\epsilon_{\rm max}$ (MeV)} \\ \cline{5-7}
                                     nucleus &&$R_{\rm box}$ (fm)&&     8  &    10  &    12  \\ \hline
\rule{0cm}{0.35cm}$^{22}$O  &&10&& 3.006 &  3.007 &  3.007 \\
                                               &&12&& 3.008 &  \underline{3.008} &  3.008  \\
                                               &&14&& 3.010 &  3.008  & 3.008 \\ \hline
                              $^{86}$Ni  &&20&& 4.535 &  4.535 & 4.535 \\
                                                &&22&& 4.544 & \underline{4.544} &  4.543 \\
                                                &&24&& 4.550 & 4.550 & 4.550 \\ \hline
                             $^{150}$Sn && 23 && 5.289 & 5.290 & 5.291 \\
                                                && 25 &&  5.294 & \underline{5.296} & 5.298 \\
                                                && 27 && 5.301 & 5.304 &  5.306 \\                     
\hline\hline
\end{tabular}
\caption{\small Neutron rms radii, expressed in fm, for the $^{22}$O, $^{86}$Ni and $^{150}$Sn
isotopes, obtained in HF(D1S)+BCS(D1S) calculations for different values 
of the parameters $R_{\rm box}$ and $\epsilon_{\rm max}$. 
The underlined values are those obtained by considering the values of 
$R_{\rm box}$ and $\epsilon_{\rm max}$ used in our standard calculations.
}
\label{tab:convergence} 
\end{center} 
\end{table} 

We have studied the numerical stability of the neutron rms radii values 
obtained in our HF+BCS calculations against 
the changes in the values of the parameters $R_{\rm box}$ and 
$\epsilon_{\rm max}$. We show
in Table \ref{tab:convergence} the values of these
radii for the nuclei  $^{22}$O, $^{86}$Ni and $^{150}$Sn
which have a large neutron excess and have been selected to be
representative of three different regions of the nuclear chart.
In the table, we have underlined the values obtained in our 
standard calculations, 
{\it i.~e.} those where we have verified the convergence of 
all the quantities obtained in BCS, such as pairing gaps, 
quasi-particle energies, occupation probabilities, etc.
The other values shown in the 
table have been obtained by varying $\rbox$ and $\emax$.
The largest relative differences with respect to the 
standard values are below 0.2\%. 
We have also checked that the neutron densities obtained for a 
given $R_{\rm box}$ and the three values of 
$\epsilon_{\rm max}$ shown in the table do not change significantly.
Henceforth, we have indicated
as HF(D1S)+BCS(D1S) the calculations carried out with  
$\epsilon_{\rm max}=10\,$MeV, and with a value of
$R_{\rm box}$, obviously different for each nucleus investigated,
but properly selected to guarantee the convergence of the results. 

In the next sections we shall compare our HF(D1S)+BCS(D1S) results
with those of a standard HFB calculation using the 
D1S interaction and collected in the public compilation of Ref. \cite{cea}. 
We have labelled these as reference results and indicated them as HFB$^{\rm ref}$(D1S).

For a further comparison, we have carried out HFB calculations with the D1S interaction 
by using the code HFBAXIAL \cite{rob02}. We have labelled HFB(D1S)
the corresponding results. In this approach, the HFB equations are solved
by making an expansion on a harmonic oscillator basis.
In this case, we have investigated the convergence of the results 
with respect to the number of expansion
terms, $N_{\rm ho}$, which plays a role analogous to that of $R_{\rm box}$ 
in our HF(D1S)+BCS(D1S) calculations. 

To analyze the effects due to the range of the interaction used in the HFB approach, 
a further benchmark comparison has been carried out with the 
results obtained with the Skyrme interaction SLy5 \cite{cha98}. For these calculations, 
which we have labelled HFB(SLy5), the code HFBRAD has been used \cite{ben05}. 
As is often the case in HFB when a zero-range effective nucleon-nucleon interaction 
is used \cite{dob84,dob96,dob96b,ben05}, the force that we have considered 
in the pairing sector is not SLy5. We have adopted a volume pairing 
field that follows the shape of the nuclear density, and we have considered the 
input variables selected for the test run in Ref. \cite{ben05}.
In HFBRAD the HFB equations are solved in
$r$-space, we have used an integration step of $0.1\,$fm, and 
we have studied  the convergence of the results with respect to the values of
$R_{\rm box}$.

\section{Results}
\label{sec:results}

\subsection{Neutron rms radii and densities of Ni isotopes}

\begin{figure}[!t] 
\begin{center} 
\includegraphics [scale=0.5,angle=90]{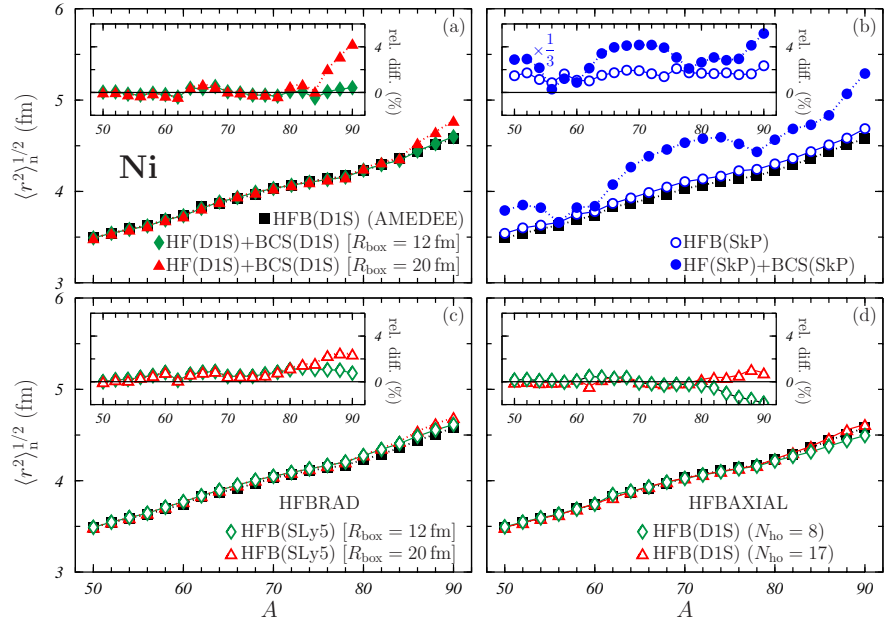} 
\vskip 0.0 cm 
\caption{\small Neutron rms radii for the Ni isotopes. 
In all the panels, the reference values taken from the AMEDEE 
database \cite{cea} are indicated by the solid black squares.
In panel (a), solid green diamonds 
and solid red triangles show the HF(D1S)+BCS(D1S) results, 
with $R_{\rm box}=12$ and $20\,$fm, respectively. 
In panel (b), open and solid blue circles indicate the results of Ref.~\cite{dob96}
corresponding, respectively, to HFB and HF+BCS calculations carried out 
with the SkP Skyrme interaction. 
In panel (c), open green diamonds and open red triangles indicate 
the results of HFB calculations carried out with the code HFBRAD~\cite{ben05}   
by using the SLy5 Skyrme interaction and
$R_{\rm box}=12$ and $20\,$fm, respectively.
In panel (d) open green diamonds and open red triangles show HFB results
obtained with the D1S interaction by using the code HFBAXIAL \cite{rob02} 
with $N_{\rm ho}=8$ and $17$, respectively. 
In the insets we show the relative differences with the reference results.
The values indicated by the solid blue circles in the inset of panel (b) 
are divided by 3 in order to fit in the scale of the figure.}
\label{fig:Rneu-Ni} 
\end{center} 
\end{figure}

Neutron rms radii of Ni isotopes with large neutron excess have 
been investigated in the past in connection with the emergence 
of the neutron gas problem \cite{dob96}.
In Fig.~\ref{fig:Rneu-Ni} we show the results obtained for  the even-even
Ni isotopes with $A$ varying from 50 to 90 ({\it i.~e.}, the neutron number $N$ varying from 22 to 62). 
In panel (a) we compare 
our HF(D1S)+BCS(D1S) results calculated by using two different values of the box size,
specifically $R_{\rm box}=12\,$fm (solid green diamonds) and $20\,$fm (solid red triangles), 
with the reference values HFB$^{\rm ref}$(D1S) (solid black squares).
In the inset, we show the corresponding relative differences, which
are well below 1\%, in absolute value, 
except for $R_{\rm box}=20$ fm and $N>58$ where they slowly increase with $N$ up to about
$4\%$ for $^{90}$Ni. A similar difference between HF+BCS and HFB calculations using the SIII Skyrme interaction was found by Grasso {\it et al.} \cite{Grasso}.

\begin{figure}[!b] 
\begin{center} 
\includegraphics [scale=0.5,angle=0]{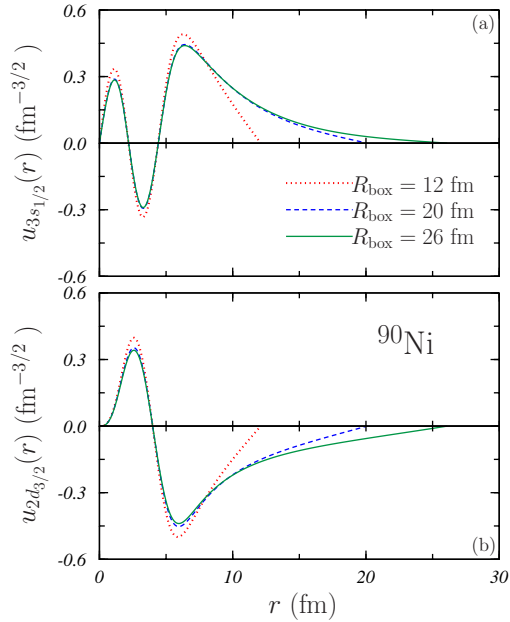} 
\vskip 0.0 cm 
\caption{\small Wave functions of the neutron $3s_{1/2}$ (a) and $2d_{3/2}$ (b) s.p. states for the $^{90}$Ni isotope obtained in our HF(D1S) calculations, with $R_{\rm box}=12$ (dotted red lines), 20 (dashed blue lines) and $26\,$fm (solid green lines).}
\label{fig:3s-2d-convergence} 
\end{center} 
\end{figure} 

The results of panel (a) should be directly compared with those of 
panel (b) where we show the neutron rms radii 
of Ref.~\cite{dob96}, obtained by using the Skyrme SkP interaction. 
In this panel, the open blue circles indicate the HFB(SkP) results and 
the solid blue circles those of the HF(SkP)+BCS(SkP) calculations. 
The relative differences between the results of Ref.~\cite{dob96} 
and the reference values (indicated as in panel (a) with the solid black squares) are shown in the inset. 
 While the HFB(SkP) results differ from the HFB$^{\rm ref}$(D1S)
by $\sim 2\%$ at most, in the case of the HF(SkP)+BCS(SkP) 
the relative  difference is above
$15\%$ for $N=62$ (or $A=90$), more than three times larger than 
our result for $R_{\rm box}=20\,$fm. 
These relative differences
are also rather large, well above 10\%, for the isotopes with $N=38-46$ ($A=66-74$)
while our HF(D1S)+BCS(D1S) results are very close to the
HFB$^{\rm ref}$(D1S) data, independently of the value of $R_{\rm box}$. 

The large disagreement between the HF(SkP)+BCS(SkP) and HFB(SkP) 
neutron rms radii shown in Fig.~\ref{fig:Rneu-Ni}(b) has been considered
as an evidence of the neutron gas problem \cite{dob96}. 
On the other hand, also the, much smaller,
increase in the neutron rms radii, starting from
the $^{86}$Ni isotope, of our HF(D1S)+BCS(D1S) 
calculations for $R_{\rm box}=20\,$fm with respect to the results found for 
$R_{\rm box}=12\,$fm, could be interpreted as indication of such 
neutron gas effects. However, to a large extent, this is due to a 
lack of numerical convergence. 
In fact, a similar situation is observed in the panel (c) of Fig.~\ref{fig:Rneu-Ni} 
where  the complete sequence of Ni neutron rms radii obtained within a 
HFB(SLy5) calculation are shown for $R_{\rm box}=12$ (open green diamonds) 
and $20\,$fm (open red triangles). In this case, for the $^{90}$Ni nucleus, 
the differences between both results and the 
reference values reached $\sim 2\%$.
However, this effect cannot be associated to a neutron gas effect
because these are HFB calculations.

Since the results shown in Fig.~\ref{fig:Rneu-Ni}(c) have been 
obtained with the zero-range SLy5 Skyrme interaction, we wondered
wether the lack of convergence pointed out above could be due to the
zero-range features of the interaction. To clarify this problem
we have carried out HFB(D1S) calculations by using the HFBAXIAL \cite{rob02} 
with $N_{\rm ho}=8$ and 17 expansion terms. 
The comparison between these results and those of Ref. \cite{cea} 
is presented in the panel (d) of Fig. \ref{fig:Rneu-Ni} with open green 
diamonds and open red triangles, respectively. 
Again, a difference of about $3\%$ is observed between the results 
of the two calculations for the heavier Ni isotopes. 
These results indicate that, 
independently of the interaction (zero or finite range) utilized, 
the theoretical approach adopted (HFB or HF+BCS) 
and the numerical method used to solve the equations
(expansion in a harmonic oscillator basis or direct 
solution of the equations in $r$ space) the 
observed behavior for nuclei with large neutron excess
is a matter of convergence.

To have a better insight on the possible emergence of neutron gas
effects in the heaviest Ni isotopes investigated, we have analyzed 
the $3s_{1/2}$ and $2d_{3/2}$ neutron s.p. states 
in $^{86}$Ni, $^{88}$Ni, and $^{90}$Ni. In our HF(D1S) calculation 
the $3s_{1/2}$ level is the last fully occupied neutron s.p. state in $^{86}$Ni, 
while the $2d_{3/2}$ is being filled up in the other two isotopes. 
The radial wave functions of these two s.p. states for $^{90}$Ni
are shown in Fig. \ref{fig:3s-2d-convergence} for 
three different values of $R_{\rm box}$. We recall that these s.p. wave functions are not changed 
by BCS. 

In both cases, the wave functions calculated with $R_{\rm box}=12\,$fm, 
represented by the dotted red lines, tend sharply to zero when $r$ approximates 
$12\,$fm in a manner that does not correspond to the physically
expected exponentially decaying tail, which, on the other hand, is more closely approached by 
the other two curves. 
We have obtained similar results for the $^{86}$Ni and $^{88}$Ni isotopes. 

The unphysical asymptotic behavior at $R_{\rm box}=12\,$fm 
is a symptom of a lack of numerical stability of the calculations. 
We present in Table \ref{tab:3s-2d-convergence} the HF(D1S) s.p.
energies, $\epsilon_{\alpha}$, and the BCS(D1S) occupation probabilities,
 $(v_\alpha^{\rm BCS})^2$, of the two neutron s.p. states that we are discussing. 
In these three Ni isotopes the $3s_{1/2}$ state
is bound, while the $2d_{3/2}$ state has a positive energy, 
{\it i.~e.} it is in the ``discretized'' continuum. 
Despite of that, the convergence of the
results is evident in both cases, though it is necessary to use, at least,
$R_{\rm box}\sim 20\,$fm to reach a reasonable numerical stability. 
These results indicate that the $2d_{3/2}$ is a quasi-bound s.p. state.

\begin{table}[!t] 
\begin{center} 
\begin{tabular}{cccccccccc}
\hline\hline
&  & \multicolumn{2}{c}{$^{86}$Ni} &~~~& \multicolumn{2}{c}{$^{88}$Ni} &~~~&  \multicolumn{2}{c}{$^{90}$Ni} \\ \cline{3-4}\cline{6-7}\cline{9-10}
s.p. state & $R_{\rm box}$ (fm) & $\epsilon_{\alpha}$ (MeV) & $(v_\alpha^{\rm BCS})^2$ && $\epsilon_{\alpha}$ (MeV) & $(v_\alpha^{\rm BCS})^2$ & & $\epsilon_{\alpha}$ (MeV) & $(v_\alpha^{\rm BCS})^2$ \\ \hline
\rule{0cm}{0.35cm}$3s_{1/2}$ & 12 & $-0.420$ & $ 0.982$ & & $-0.585$ & $ 0.962$ & & $-0.786$ & $ 1.000$ \\
                                                 & 20 & $-0.668$ & $ 1.000$ & & $-0.775$ & $ 0.989$ & & $-0.872$ & $ 1.000$ \\
                                                 & 26 & $-0.673$ & $ 0.999$ & & $-0.766$ & $ 0.990$ & & $-0.864$ & $ 0.999$ \\ \hline
\rule{0cm}{0.35cm}$2d_{3/2}$ & 12 & $ 0.834$ & $ 0.010$ & & $ 0.554$ & $ 0.494$ & & $ 0.265$ & $ 1.000$ \\
                                                & 20 & $ 0.546$ & $ 0.000$ & & $ 0.371$ & $ 0.501$ & & $ 0.200$ & $ 1.000$ \\
                                                & 26 & $ 0.502$ & $ 0.001$ & & $ 0.363$ & $ 0.494$ & & $ 0.204$ & $ 0.981$ \\
\hline\hline
\end{tabular}
\caption{\small HF s.p. energies, $\epsilon_{\alpha}$, and BCS occupation probabilities, $(v_\alpha^{\rm BCS})^2$, of the $3s_{1/2}$ and $2d_{3/2}$ neutron s.p. states for the $^{86}$Ni, $^{88}$Ni and $^{90}$Ni isotopes, calculated with the D1S interaction for various values of the parameter $R_{\rm box}$. In these calculations we have considered $\epsilon_{\rm max}=10\,$MeV. }
\label{tab:3s-2d-convergence} 
\end{center} 
\end{table} 

\begin{figure}[!b] 
\begin{center} 
\includegraphics [scale=0.5,angle=0]{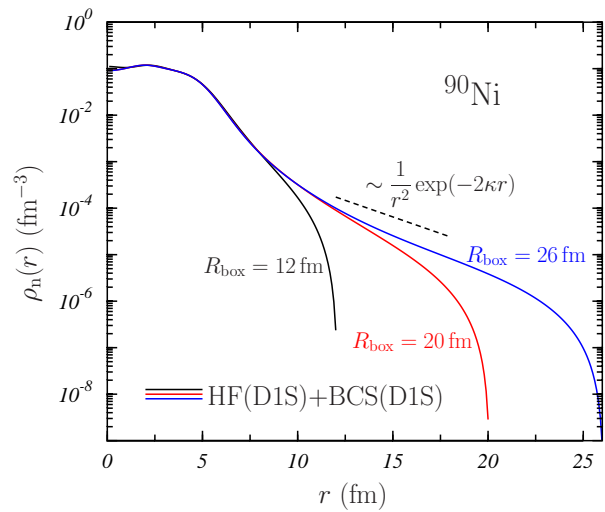} 
\vskip 0.0 cm 
\caption{\small Neutron density distributions for the $^{90}$Ni nucleus obtained in our 
HF(D1S)+BCS(D1S) calculations. Solid black, red and blue curves correspond, respectively,
to $R_{\rm box}=12$, 20, and $26\,$fm. In all these calculations we have chosen 
$\epsilon_{\rm max}=10\,$MeV. The dashed straight line indicates the expected 
 behavior of the density, according to Eq.~(\ref{eq:asymp}). }
\label{fig:90Ni-density} 
\end{center} 
\end{figure} 

In Fig.~\ref{fig:90Ni-density} we show the neutron density of the 
$^{90}$Ni isotope calculated with HF(D1S)+BCS(D1S) for 
$R_{\rm box}$=12 (black curve), 20 (red curve) and 26 fm 
(blue curve). The dashed line indicates the expected 
asymptotic behavior
\beq
\rho_{\rm n}(r) \, \sim \, \displaystyle  \frac{1}{r^2}\exp(-2\kappa r) \, ,
\label{eq:asymp}
\eeq
where $\kappa=\sqrt{|E_{\rm min}-\lambda| 2 m_{\rm n}/\hbar^2}$.
Here $E_{\rm min}=0.330\,$MeV  is the value
of the quasi-particle energy of the state defining the Fermi level
for neutrons, the $2d_{3/2}$ in this case, and $\lambda=0.522\,$MeV 
is the BCS chemical potential. As it can be seen, this behavior is well reproduced
at large $r$-values and no evidence of neutron gas effects is found.

\subsection{Neutron rms radii and densities of Sn isotopes}

Several Sn isotopes have been studied in the literature in connection with the neutron 
gas problem: abnormally high tails of their neutron densities \cite{dob96b} 
and huge growths of the neutron rms radii \cite{dob84,ono10} 
were considered as its signature. 

We present in Table \ref{tab:Sn-rms}
the values of the rms radii of the neutron distributions 
corresponding to various even-even Sn isotopes, focussing on the 
convergence of the results in the three types of calculations that 
we have carried out.
We have selected those Sn isotopes with spherical shape.
The only exceptions are the $^{150}$Sn and $^{160}$Sn 
nuclei for which the HFB calculations of Ref. \cite{cea} indicate
a small deformation. We have included them in our investigation 
because they have been specifically considered 
in the study of the neutron gas problem \cite{dob96b}.
In any case, we have carried out deformed HFB(D1S) calculations 
with the HFBAXIAL code and 
we have verified that the neutron rms radii values obtained for 
these two nuclei differ from those found in the 
spherical case by less than $0.04\%$. 

\begin{table}[htb] 
\begin{center} 
\begin{tabular}{ccccccccccccccccccccc}
\hline\hline
 & \rule{0.5cm}{0cm} & \multicolumn{4}{c}{HF(D1S)+BCS(D1S)} & \rule{0.5cm}{0cm} & \multicolumn{4}{c}{HFB(D1S)} & \rule{0.5cm}{0cm} & \multicolumn{4}{c}{HFB(SLy5)}\\ \cline{3-6}\cline{8-11}\cline{13-16}
&&\multicolumn{4}{c}{$R_{\rm box}$ (fm)} && \multicolumn{4}{c}{$N_{\rm ho}$} &&\multicolumn{4}{c}{$R_{\rm box}$ (fm)} \\ \cline{3-6}\cline{8-11}\cline{13-16}
                                     nucleus && 10 & 15 & 20 & 25 && 8 &11 & 14 & 17 && 10 & 15 & 20 & 25  \\ \hline
\rule{0cm}{0.35cm}$^{140}$Sn  && 4.93 & 4.98 & 4.98 & 4.98 && 4.97 & 4.97 & 4.99 & 4.99 && 5.00 & 5.05 & 5.05 & 5.05 \\
                              $^{142}$Sn  && 4.96 & 5.04 & 5.06 & 5.07 && 5.00 & 5.01 & 5.03 & 5.04 && 5.04 & 5.09 & 5.09 & 5.09 \\
                              $^{144}$Sn  && 5.02 & 5.04 & 5.04 & 5.04 && 5.04 & 5.05 & 5.07 & 5.08 && 5.07 & 5.13 & 5.13 & 5.13 \\
                              $^{146}$Sn  && 5.02 & 5.07 & 5.07 & 5.07 && 5.07 & 5.09 & 5.11 & 5.12 && 5.10 & 5.17 & 5.18 & 5.18 \\
                              $^{150}$Sn  && 5.10 & 5.24 & 5.28 & 5.30 && 5.14 & 5.16 & 5.18 & 5.20 && 5.17 & 5.25 & 5.25 & 5.26 \\ 
                              $^{160}$Sn  && 5.27 & 5.41 & 5.44 & 5.44 && 5.27 & 5.30 & 5.33 & 5.35 && 5.30 & 5.42 & 5.43 & 5.43\\
                              $^{164}$Sn  && 5.31 & 5.45 & 5.48 & 5.49 && 5.31 & 5.35 & 5.38 & 5.41 && 5.35 & 5.47 & 5.49 & 5.49\\
                              $^{166}$Sn  && 5.32 & 5.47 & 5.49 & 5.50 && 5.33 & 5.38 & 5.41 & 5.43 && 5.38 & 5.50 & 5.51 & 5.51\\
                              $^{168}$Sn  && 5.34 & 5.48 & 5.50 & 5.51 && 5.35 & 5.40 & 5.43 & 5.45 && 5.40 & 5.52 & 5.53 & 5.53 \\
                              $^{170}$Sn  && 5.35 & 5.49 & 5.51 & 5.52 && 5.37 & 5.42 & 5.45 & 5.47 && 5.42 & 5.54 & 5.55 & 5.55\\
                              $^{172}$Sn  && 5.36 & 5.50 & 5.52 & 5.52 && 5.39 & 5.44 & 5.47 & 5.49 && 5.44 & 5.56 & 5.57 & 5.57 \\
\hline\hline
\end{tabular}
\caption{\small 
Neutron rms radii for various Sn isotopes calculated with
HF(D1S)+BCS(D1S), with HFB(D1S), by using the HFBAXIAL code, 
and with HFB(SLy5), by using the HFBRAD code, approaches. 
The convergence with the corresponding values of
$R_{\rm box}$ or $N_{\rm ho}$ is analyzed. 
The HF(D1S)+BCS(D1S) calculations have been carried out
with $\epsilon_{\rm max}=10\,$MeV.
}
\label{tab:Sn-rms} 
\end{center} 
\end{table} 

The results of Table \ref{tab:Sn-rms} indicate the numerical 
stability of all our calculations for $\rbox$ values larger than
20 fm  or $N_{\rm ho}$ larger than 14. In addition, 
the results obtained in the three approaches show a good agreement, 
with relative differences smaller than $2\%$. 
We do not observe any anomaly related to our
HF(D1S)+BCS(D1S) results, which again do not show neutron gas effects. 

Our results disagree with those of Ref.~\cite{dob84} where it is shown
that the neutron rms radii of the $^{120}$Sn and $^{160}$Sn nuclei
grow with $R_{\rm box}$ in HF+BCS calculations, 
while remain almost constant in HFB.
The relative differences between the results of the two calculations, 
carried out with $R_{\rm box}=25\,$fm, are 
about  6\% and 20\%, respectively, for the two Sn isotopes. 
Similar results are presented in Ref.~\cite{ono10} where it is
shown that the values of the neutron rms radii obtained 
in the HF+BCS approach grow with the increasing number 
of oscillator shells $N_{\rm ho}$ used in the calculation. 

\begin{figure}[ht] 
\begin{center} 
\includegraphics [scale=0.5,angle=0]{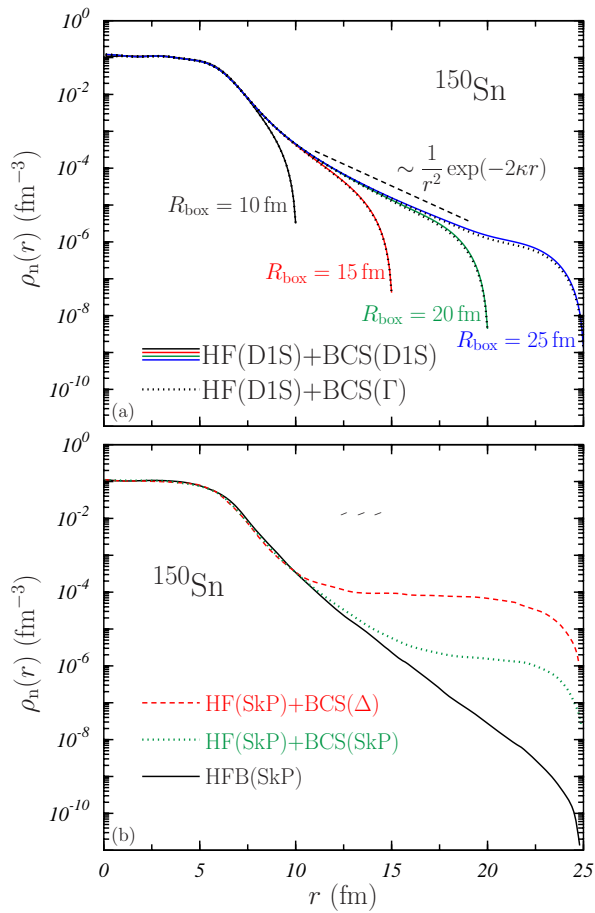} 
\vskip 0.0 cm 
\caption{\small Neutron density distributions for the $^{150}$Sn nucleus. 
In panel (a), the densities obtained in our HF(D1S)+BCS(D1S) calculations 
are shown by solid curves for $R_{\rm box}=10$ (black), 
15 (red), 20 (green) and $25\,$fm (blue). The dotted black curves 
indicate the HF(D1S)+BCS($\Gamma$) results
obtained for various $R_{\rm box}$ values. 
In all these calculations we have used $\epsilon_{\rm max}=10\,$MeV. 
The dashed straight line indicates the expected asymptotic behavior 
of the density, according to Eq.~(\ref{eq:asymp}). 
In panel (b), we show the densities presented in Figs.~14 and 19 
of Ref.~\cite{dob96b}.
The HFB(SkP) density (solid black curve) is compared to 
those obtained in HF(SkP)+BCS($\Delta$) (dashed red curve) 
and HF(SkP)+BCS(SkP) (dotted green curve) calculations.
}
\label{fig:150Sn-density} 
\end{center} 
\end{figure} 

\begin{figure}[!t] 
\begin{center} 
\includegraphics [scale=0.5,angle=0]{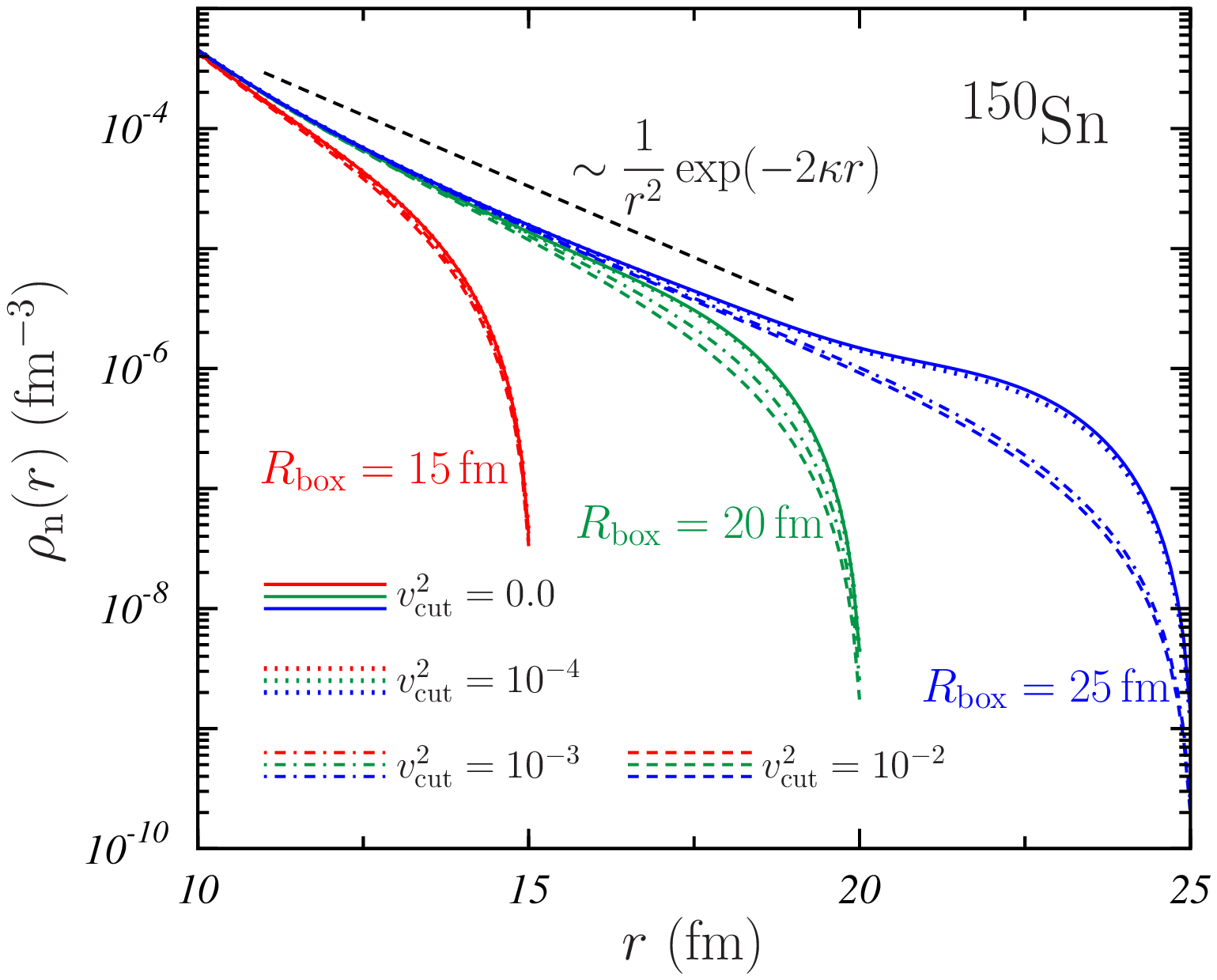} 
\vskip 0.0 cm 
\caption{\small Neutron density distributions for the $^{150}$Sn nucleus obtained in 
our HF(D1S)+BCS(D1S) calculations for $R_{\rm box}=15$ (red lines), 
20 (green lines) and $25\,$fm (blue lines) by using $\epsilon_{\rm max}=10\,$MeV. 
Solid, dotted, dashed-dotted and dashed curves 
include all s.p. states with occupation probabilities $(v_\alpha^{\rm BCS})^2$ larger than
0, $10^{-4}$, $10^{-3}$ and $10^{-2}$, respectively. 
The dashed straight line indicates the expected asymptotic behavior 
of the density, according to Eq.~(\ref{eq:asymp}). 
}
\label{fig:vcut-density} 
\end{center} 
\end{figure} 

We have analyzed  the neutron density distributions
of various Sn isotopes. As example of this investigation,
we present in Fig.~\ref{fig:150Sn-density} the results obtained 
for $^{150}$Sn, one of the isotopes studied in Ref.~\cite{dob96b} 
to show the appearance of the neutron gas problem.
The full lines in the panel (a) indicate the results 
of our HF(D1S)+BCS(D1S) calculations for different values of 
$R_{\rm box}$ from 10 to $25\,$fm. 
The dashed straight line shows the expected 
asymptotic behavior defined in Eq. (\ref{eq:asymp}),
with $E_{\rm min}=0.561\,$MeV, corresponding 
to the neutron $3p_{1/2}$ s.p. level, and $\lambda=-0.304\,$MeV. 
This behavior is approximately reproduced
at large $r$-values, independently of the $\rbox$ value. 
Our density distributions do not show any neutron gas effect. 

The curves shown in panel (b) of Fig. \ref{fig:150Sn-density} 
have been adapted from Figs. 14 and 19 of Ref. \cite{dob96b}. 
They correspond to calculations that were carried out by using 
the Skyrme SkP interaction \cite{dob84}. In comparison with the 
HFB(SkP) density (solid black line), the HF+BCS distributions 
(dashed red and dotted green curves) show large asymptotic 
tails that are extremely sensitive to the interaction used for the pairing field.
The dotted green curve shows the result obtained when the
SkP interaction is used consistently in both the HF and BCS 
parts of the calculation. If in the pairing channel the SkP force is substituted 
by a contact interaction with strength $\Delta$, the tail of the 
density distribution increases by more than one order 
of magnitude, as it is shown by the dashed red curve. 

The results of Refs.~\cite{dob84,dob96b} indicate a large sensitivity 
of the tails of the neutron density distributions to the interaction used in the BCS calculations.
We have investigated if our results also show the same kind of sensitivity and if the neutron 
gas effects could appear when a zero-range interaction is utilized instead of a finite-range force.
To clarify this point we have carried out BCS calculations with 
a $\delta-$contact interaction of the type
\begin{equation}
V({\bf r},{\bf r}') \, = \, \Gamma \, \delta({\bf r}-{\bf r}') 
\label{eq:contact}
\end{equation}
together with the HF(D1S) set of s.p. energies and wave functions. 
We have labelled HF(D1S)+BCS($\Gamma$) the corresponding results. 
The value of $\emax$ is 
the same adopted in our usual HF(D1S)+BCS(D1S) calculations. 
The strength $\Gamma$ in Eq. (\ref{eq:contact})
has been chosen to reproduce the average HF(D1S)+BCS(D1S) gap value,
\begin{equation}
\overline{\Delta} \, = \,\displaystyle \frac{\displaystyle\sum_\alpha \, (v_\alpha^{\rm BCS})^2 \, \Delta_\alpha}{\displaystyle\sum_\alpha  \, (v_\alpha^{\rm BCS})^2} \, .
\label{eq:avgap}
\end{equation}
In the previous equation $\Delta_\alpha$ indicates the BCS gap 
of the corresponding quasiparticle state.

The results of these calculations are shown in the panel (a) 
of Fig. \ref{fig:150Sn-density} by the dotted black curves.
These density distributions are very similar to 
the HF(D1S)+BCS(D1S) ones and do not show
any indication of the presence of the neutron gas.  
This is also confirmed by the fact that the values of the 
neutron rms radii differ from those given in Table \ref{tab:Sn-rms} 
by less than 0.1\% , at maximum, which is the case of $R_{\rm box}=25\,$fm.
We have obtained similar results for all the Sn isotopes investigated. 

A careful observation of Fig. \ref{fig:150Sn-density}(a) 
indicates that the density obtained with $R_\textrm{box}=25\,$fm 
has a small hump in the region above $20\,$fm. 
To understand the source of this behavior,
we have studied the relative importance of the s.p. states contributing
to this density distribution, {\it i.~e.} to the diagonal part of the 
OBDM defined in Eq. (\ref{eq:obdmbcs}). 
As each s.p. wave function is weighted by 
its occupation probability $(v^\textrm{BCS}_\alpha)^2$,
we have surmised that the hump could be generated by the presence of s.p.
states with very small occupation probabilities.
For this reason, we have calculated the density distribution by selecting
s.p. states with $(v^\textrm{BCS}_\alpha)^2 > v_{\rm cut}^2$.

The results of this study are presented in  Fig. \ref{fig:vcut-density}
where the densities obtained with $v_{\rm cut}^2=10^{-4}$ (dotted lines),
$v_{\rm cut}^2=10^{-3}$ (dashed dotted lines) and $v_{\rm cut}^2=10^{-2}$
(dashed lines) are compared with those of 
Fig~\ref{fig:150Sn-density}(a), which correspond to $v_{\rm cut}^2=0$ (full lines).
These results indicate that the hump at large $r$
values is clearly due to the contribution of s.p. states with very small
occupation probabilities. The effect becomes larger with increasing
$R_\textrm{box}$, as it is evident for the results corresponding to
$R_\textrm{box}=25\,$fm. It is also worth pointing out that,
in this case, the densities obtained with $v_{\rm cut}^2=10^{-3}$ 
or $10^{-2}$ verify much better the expected asymptotic behavior 
indicated in the figure by the black dashed straight line. 

We have further investigated the relevance of the s.p. states with 
small occupation
probabilities, by comparing our HF(D1S)+BCS(D1S) calculations 
for different values of $v_{\rm cut}^2$
with the HFB(SLy5) results obtained by using the HFBRAD code.
In order to make a fair comparison we have 
renormalized our HF(D1S)+BCS(D1S) density distributions to the 
correct number of nucleons. 
In all cases, $R_\textrm{box}=25\,$fm has been considered.

\begin{figure}[!t] 
\begin{center} 
\includegraphics [scale=0.5,angle=0]{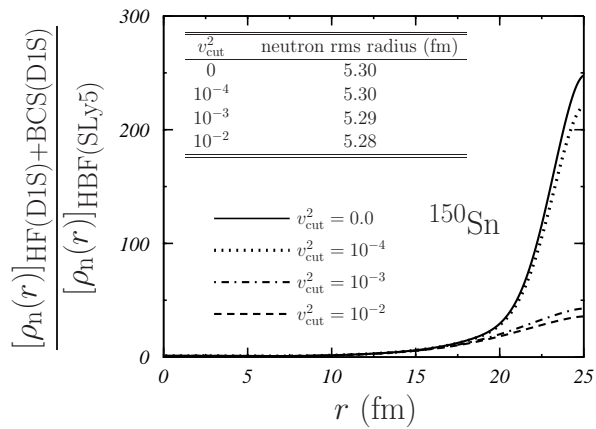} 
\vskip 0.0 cm 
\caption{\small 
Ratios between the neutron densities generated by 
our HF(D1S)+BCS(D1S) calculations, with
$R_{\rm box}=25$ and $\epsilon_{\rm max}=10\,$MeV, 
and that obtained in a HFB(SLy5) calculation carried out with HFBRAD. 
Solid, dotted, dashed-dotted and dashed curves include, 
in the HF+BCS calculations, all s.p. states with occupation probabilities 
$(v_\alpha^{\rm BCS})^2$ larger than 0, $10^{-4}$, $10^{-3}$ and $10^{-2}$, respectively. 
The neutron rms radii obtained in the HF(D1S)+BCS(D1S) calculations 
are given in the inset. 
}
\label{fig:density-rat} 
\end{center} 
\end{figure} 

In Fig.~\ref{fig:density-rat} we show the ratio between the 
HF(D1S)+BCS(D1S) densities, 
obtained by using the prescription above described for different
values of $v_{\rm cut}^2$, and that of the HFB(SLy5) calculation.
These ratios remain close to 1 up to about $15\,$fm and grow for larger $r$ values. 
The strong reduction of the ratio observed for $r\geq 20\,$fm when 
$v^2_{\rm cut}=10^{-2}$ and $10^{-3}$, 
shows the relevance of s.p. states with small occupation probability.
In the inset of the figure, we also present the values of 
the corresponding neutron rms radii, which coincide with those 
obtained before the renormalisation of the densities. The relative difference 
among these values is $0.2\%$ at most, similar to the differences found between 
the neutron rms radii shown in Table \ref{tab:Sn-rms}. It is worth mentioning that the 
results obtained for the rms neutron radii of some Sn isotopes by Del Estal {\it et al.} in a BCS calculation, 
considering all s.p. states with positive energies or just those 
corresponding to quasi-bound states show the same relative differences \cite{Estal}.

\subsection{Proton elastic cross sections}

The results we have discussed so far indicate that our
HF+BCS calculations do not show the large neutron gas
effects mentioned in Refs. \cite{dob84,dob96,dob96b,dob99,ono10}.
On the other hand, we have found that at large $r$ values, s.p.
states with small occupation probability slightly modify the
expected behavior of the density distributions. Slightly means
that these modifications are several order of magnitude smaller
than the maximum values of the distributions, and in the figures
we emphasize them by using a logarithmic scale. 
In this section we investigate the relevance of these modifications
of the neutron density distributions at large $r$ values in the
calculations of other observables.

A first answer to this question is already provided by the 
results of Table \ref{tab:Sn-rms} showing that 
the neutron rms radii are extremely stable against the
increase of the $R_\textrm{box}$ values. 
This means that the s.p. states
generating the small hump observed at large $r$ 
are irrelevant in the calculation of these radii. 

We have also studied the possible effects of these states on 
the differential cross section for proton elastic scattering off $^{150}$Sn.
We have chosen this nucleus since, in 
Ref. \cite{dob96b}, shows large neutron gas effects. 
The calculations of the cross sections have been done by adopting 
a rather simple optical potential generated by folding 
the corresponding total matter density, {\it i.~e.} the sum of 
the proton and neutron densities, 
with a contact nucleon-nucleon potential:
\begin{equation}
V({\bf r},{\bf r}') \, = \, -V_0 \, \delta({\bf r}-{\bf r}') \, .
\label{eq:deltaf}
\end{equation}
We have chosen the strength value as $V_0=50\,$MeV.

\begin{figure}[!ht] 
\begin{center} 
\includegraphics [scale=0.5,angle=90]{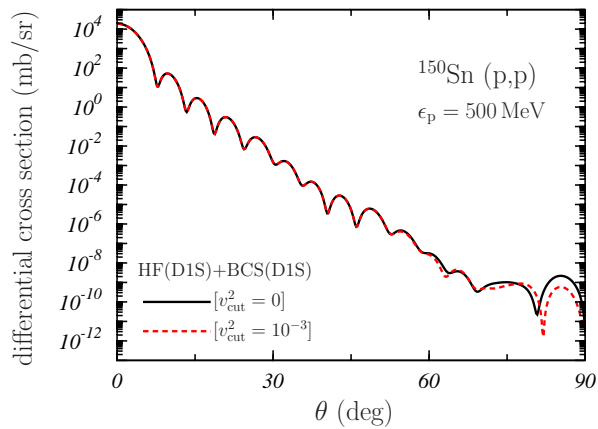} 
\vskip 0.0 cm 
\caption{\small 
Differential cross sections for elastic scattering of $500\,$MeV protons off 
$^{150}$Sn nucleus. The optical potential used in the calculations
has been obtained by folding the 
nucleon-nucleon potential defined in Eq. (\ref{eq:deltaf}) with 
the matter densities generated in our
HF(D1S)+BCS(D1S) approach, with $R_{\rm box}=25\,$fm 
and $\epsilon_{\rm max}=10\,$MeV. 
The solid black curve has been obtained with a matter density 
calculated by considering all the BCS s.p. states, 
while the dashed red curve shows the results obtained
by using a, renormalized, matter density where only the s.p. states with 
BCS occupation $(v_\alpha^{\rm BCS})^2  > 10^{-3}$ have been taken
into account.
}
\label{fig:cross} 
\end{center} 
\end{figure} 

The results obtained are shown in Fig. \ref{fig:cross} for a proton 
incident energy of $500\,$MeV. In this calculation the matter densities 
have been obtained by using the s.p. wave functions obtained in our
HF(D1S)+BCS(D1S) with $\rbox=25\,$fm and $\emax=10\,$MeV.
The solid black line indicates the result obtained by considering the
full set of BCS s.p. wave functions, while the dashed red curve that
obtained by considering $v^2_\textrm{cut} > 10^{-3}$, and, of course,
properly renormalizing the density.  
The two curves are almost overlapping up to a scattering angle 
of about $60\,$degrees. Some differences occur above this value where 
the differential cross section is 14 order of magnitude smaller than 
the maximum obtained at $\theta=0$. We have checked that, for incident energies smaller 
than 400 MeV, the corresponding curves overlap up to $90\,$degrees.

\section{Summary and conclusions}
\label{sec:conclusions}

From the theoretical point of view,
BCS calculations are affected by neutron gas effects that
are instead absent, by construction, in the HFB approach. 
These effects are due to the s.p. states with positive energies and 
we have studied their quantitative relevance 
by comparing the results of HF+BCS and HFB calculations
carried out with various interactions and different numerical techniques. 

We have not found any remarkable evidence of the neutron gas problem in all the 
observables analyzed. The contribution of the largest part of the s.p. states 
with positive energy is quantitatively irrelevant due to the fact that their occupation 
probabilities are very small.

We have also excluded that the use of a zero-range interaction in the 
pairing sector could enhance the effect of the neutron gas. 

We have pointed out that a proper choice of the input parameters 
ensures a good numerical convergence of the values of the neutron rms radii.

We have investigated the role of the s.p. states with small occupation 
probabilities in the neutron density distributions 
and we have found that they produce a
small hump at large distances from the nuclear center. This effect is due to accumulated contributions 
of a large number of these s.p. states.
This enhancement in the density is, however, orders of 
magnitude smaller than those claimed 
in the literature as a neutron gas evidence.

These results indicate the reliability of the HF+BCS approach in
the description of nuclei with large neutron excess.

\acknowledgments 
The authors acknowledge F. Salvat for providing us with the code to calculate the proton cross sections. 
This work has been partially supported by  
the Junta de Andaluc\'{\i}a (FQM387), the Spanish Ministerio de 
Econom\'{\i}a y Competitividad (FPA2015-67694-P) and the European 
Regional Development Fund (ERDF). 
 
%

\end{document}